\documentclass[compsoc,conference,a4paper,10pt,times]{IEEEtran}
\IEEEoverridecommandlockouts
\UseRawInputEncoding
\usepackage{booktabs}
\usepackage{cite}
\usepackage{amsmath,amssymb,amsfonts}
\usepackage{algorithmic}
\usepackage{graphicx}
\usepackage{textcomp}
\usepackage{bmpsize}
\usepackage{xcolor}
\usepackage{lipsum}
\usepackage[colorlinks=true,urlcolor=black,breaklinks]{hyperref}
\usepackage{multirow}
\usepackage{enumitem}
\usepackage{makecell}
\usepackage{url}
\usepackage{textcomp}

\newcommand{\comment}[1]{}
\makeatletter
\newcommand{\linebreakand}{%
  \end{@IEEEauthorhalign}
  \hfill\mbox{}\par
  \mbox{}\hfill\begin{@IEEEauthorhalign}
}
\makeatother

\def\BibTeX{{\rm B\kern-.05em{\sc i\kern-.025em b}\kern-.08em
    T\kern-.1667em\lower.7ex\hbox{E}\kern-.125emX}}
\begin{document}


\title{No Time for Downtime: Understanding Post-Attack Behaviors by Customers of Managed DNS Providers
\thanks{
This work is part of the NWO: MASCOT project, which is funded by the Netherlands Organization for Scientific Research (CS.014).
This material is based on research sponsored by the National Science Foundation (NSF) grant OAC-1724853 and OAC-2131987. The views and conclusions contained herein are those of the authors and should not be interpreted as necessarily representing the official policies or endorsements, either expressed or implied, of NSF.} }

\author{
\IEEEauthorblockN{\hspace{-0.7cm}Muhammad Yasir Muzayan Haq}
\IEEEauthorblockA{
\textit{\hspace{-0.7cm}University of Twente}\\
\hspace{-0.7cm}Enschede, The Netherlands \\
\hspace{-0.7cm}m.y.m.haq@utwente.nl}
\and
\IEEEauthorblockN{\hspace{-0.7cm}Mattijs Jonker}
\IEEEauthorblockA{
\textit{\hspace{-0.7cm}University of Twente}\\
\hspace{-0.7cm}Enschede, The Netherlands \\
\hspace{-0.7cm}m.jonker@utwente.nl}
\and
\IEEEauthorblockN{Roland van Rijswijk-Deij}
\IEEEauthorblockA{
\textit{University of Twente}\\
Enschede, The Netherlands\\
r.m.vanrijswijk@utwente.nl}
\linebreakand
\IEEEauthorblockN{kc claffy}
\IEEEauthorblockA{
\textit{CAIDA / UC San Diego}\\
La Jolla, CA, USA \\
kc@caida.org}
\and
\IEEEauthorblockN{Lambert J.M. Nieuwenhuis}
\IEEEauthorblockA{
\textit{University of Twente}\\
Enschede, The Netherlands\\
l.j.m.nieuwenhuis@utwente.nl}
\and
\IEEEauthorblockN{Abhishta Abhishta}
\IEEEauthorblockA{
\textit{University of Twente}\\
Enschede, The Netherlands\\
s.abhishta@utwente.nl}
}

\maketitle

\begin{abstract}


We leverage large-scale DNS measurement data on authoritative name servers to study the reactions of domain owners affected by the 2016 DDoS attack on Dyn.
We use industry sources of information about domain names to study the influence of factors such as industry sector and website popularity on the willingness of domain managers to invest in high availability of online services. Specifically, we correlate business characteristics of domain owners with their resilience strategies in the wake of DoS attacks affecting their domains.  Our analysis revealed correlations between two properties of domains -- industry sector and popularity -- and post-attack strategies.  Specifically, owners of more popular domains were more likely to re-act to increase the diversity of their authoritative DNS service for their domains.  Similarly, domains in certain industry sectors were more likely to seek out such diversity in their DNS service.
For example, domains categorized as \textit{General News} were nearly 6 times more likely to re-act than domains categorized as \textit{Internet Services}.
Our results can inform managed DNS and other network service providers regarding the potential impact of downtime on their customer portfolio. 
\end{abstract}


\section{Introduction}

The domain name system (DNS) is a core component of the Internet. Its primary function is to translate human-readable domain names to IP addresses. As such, e-commerce with a web site is no longer possible if the authoritative name servers for the web site's domain name fail to function. A managed DNS service provider offers premium services to manage DNS traffic~\cite{abhishta_measuring_2019}, improving reliability and performance, including potentially offering DDoS protection. 
Failures of managed DNS services adversely affect parties that rely on these services~\cite{bolstridge2016dyn}, depending on their degree of reliance.


Businesses that use a managed DNS service may adopt a more resilient strategy, i.e., use multiple DNS service providers, that ensures higher availability even when that managed DNS service provider may be unavailable. Previous studies  \cite{abhishta_measuring_2019,bates2018evidence} have shown that when services of Dyn, a large managed DNS service provider, became unavailable in 2016 \cite{bolstridge2016dyn}, many but not most domains adopted this more resilient strategy to counter the effects of downtime. 
Some businesses value continuous online availability enough to pay the premium charged by managed DNS service providers. It depends on the expected return on investment of such high availability of their online services. Different businesses value high  availability differently, but outages clearly sometimes directly induce a loss of revenue~\cite{akamai2018}.



We perform an analysis of large-scale longitudinal active DNS measurement data to identify domains that responded to the Dyn DDoS-induced outage. We map these domains to their popularity rank using historical Alexa rankings. We also obtain industry-sourced data on industry sector and risk characteristics using McAfee's URL categorization service. 
\noindent Overall, we make the following contributions:
\begin{itemize}
    
    \item We characterize 106K domains under \texttt{.com}, \texttt{.net}, and \texttt{.org} affected by the Dyn outage based on four aspects: post-attack resilience strategy, popularity, industry sector, and risk characteristics.
    
    \item 
    We observe that domains with higher popularity (based on their Alexa ranks), more commonly adopted a more resilient strategy. We find that nearly 40\% of Dyn customers in the top 100 of Alexa 1M list re-acted after the attack while in the top 500K only $\approx$15\% re-acted.
    
    \item 
    Some industry sectors had a more significant correlation with post-attack response of domain managers. For example, the probability of a domain owner to adopt a more resilient strategy after the attack was nearly 9 times more if the domain belonged to \textit{General News} category (compared to categories which did not re-act). Some industry sectors (such as \textit{Religion}) did not have a statistically significant impact on the decision of domain owners.

    \item 
    We show differences in the post-attack responses across domains categorized by their risk characteristics, based on McAfee's URL categorization service. Nearly, 9\% of the domains categorized as malicious used an additional DNS service provider after the attack, while only 1\% of domains categorized as low risk took this strategy.
\end{itemize}

\noindent Our results empirically show that factors such as industry sector and popularity of domain name influence the resilience strategy adopted by domain managers in the face of unplanned unavailability of managed DNS services.
We believe that our results will be useful for managed DNS providers in understanding the possible aftermath of unforeseen downtime. 

We first review related work (\S\ref{sec:related_work}), and then
present our data sets and analysis  methods (\S\ref{sec:methodology}), 
results (\S\ref{sec:Result_and_Discussion}), and discuss 
our conclusions and implications (\S\ref{sec:conclusion}). 

We use mostly open source data-sets: Alexa top 1M historical archives from TU M{\"u}nchen\footnote{\url{https://toplists.net.in.tum.de/archive/alexa/}} 
and URL categorization portal from McAfee~\cite{noauthor_mcafee_trustedsource}.
OpenINTEL data are available to academic researchers upon request~\cite{OpenINTEL-JSAC2016}. We use \texttt{statsmodels} package~\cite{seabold2010statsmodels} in python to implement logistic regression.

\section{Related Work}
\label{sec:related_work}

Earlier work has shown that even large-scale DNS deployments, such as the root zone of the DNS, can be a victim of DDoS attacks. Moura et al.\ provided a detailed analysis of such an attack on the root in 2015, which not only affected operations of one root server family, but led to collateral damage to other DNS operators~\cite{MouraIMC2016}.

Allman~\cite{allman_comments_2018} reflected on the structural robustness of the DNS, and provided recommendations to improve operator practices, including a call to action to implement topological and geographical diversity in terms of authoritative name server deployment. Moura et al.\ studied how DNS caching and retry behavior by resolvers may hide up to 90\% of DDoS attacks on the DNS from the perspective of Internet users, but also noted that this resolver behavior may impose a significant additional load on servers under attack~\cite{MouraIMC2018}.


%

Centralization of DNS services and resolution is increasing -- up to 30\% of queries to two ccTLDs originated from just five cloud providers in 2020~\cite{MouraIMC2020}. Such
centralization enables economies of scale in
provisioning infrastructure, but can also
present a single point of failure~\cite{bates2018evidence}.



The technical and business implications of the attack on Dyn were discussed extensively\cite{antonakakis2017understanding,bates2018evidence}. Antonakakis et al.\ analyzed the emergence of Mirai, the botnet used for attacking Dyn, while also presenting the possibility that the attackers were targeting not only Dyn but also Sony Playstation infrastructure \cite{antonakakis2017understanding}. Bates et al.\ analyzed the market share of Dyn within the top 1000 Alexa rank domains (that belonged to \texttt{.com}/\texttt{.net}/\texttt{.org} TLDs) and found that its market share dropped from 10\% before the attack to 7.5\% within the month after the attack \cite{bates2018evidence}. 

In this paper, we extend on previous work in which we analyzed 
the immediate impact of the attack on Dyn, and an earlier attack on another managed DNS provider, NS1~\cite{abhishta_measuring_2019}. In the previous paper, we showed that in terms of risk management, using multiple DNS providers is a good strategy. In this paper, we extend and analyze factors that potentially drove the decision-making of Dyn customers and characterize the behavior of different types of customers when faced with the consequences of such a large-scale attack.





\section{Data Sources and Methodology}
\label{sec:methodology}

We outline the data sources that we use in this study, and we present our methodology in detail.

\subsection{Data Sources}
We primarily used three data sources for this study: longitudinal DNS measurement data, website popularity rankings, and URL categorization data.


\textbf{Longitudinal DNS measurement data:\ } We obtained data from the OpenINTEL~\cite{OpenINTEL-JSAC2016} project, which measures a large part of the global DNS namespace on a daily basis. OpenINTEL collects, through active querying, resource records configured for domain names. The collected data include: (1) mappings of domain names to authoritative name server records; (2) encountered CNAME records along with their expansion; and (3) mappings of domain names (including name server names) to IP addresses.
Using these data points, we indicate the authoritative name servers of each domain name.
%
We consider domain names under \texttt{.com}, \texttt{.net}, and \texttt{.org} that used Dyn at any point in time on one day prior to the Dyn attack (Oct. 20, 2016) and monitor them for 20 days following the attack (Nov. 10, 2016). We also identify the use of non-Dyn servers by the same names.


\textbf{Website popularity ranking:\ }We measure domain popularity on the basis of the Alexa Top 1 million website list \cite{noauthor_alexa_nodate}. 
We extract the daily rank of each domain name from the Alexa Top 1M list for 90 days prior to the incident and calculate the median value. We create six cumulative groups of domains based on median rank, namely: Top 100, Top 1K, Top 10K, Top 100K, Top 500K, and Top 1M. For example, the group Top 1K consists of domains in the dataset with a median rank between 1 and 1K, inclusive.

\textbf{URL Categorization:\ } For our analysis, we use a URL categorization service from McAfee~\cite{noauthor_mcafee_trustedsource}. This service uses human-supervised machine learning to classify domain names based on security-related features as well as hosted content (where present).\footnote{We also considered URL categorization services from other providers i.e., FortiGuard and Zvelo. However, FortiGuard only provides a single category label for each domain name while we believe that a domain name might belong to multiple categories such as \texttt{microsoft.com} (can be \textit{Business}, \textit{Software}, and \textit{Hardware}).
Meanwhile, Zvelo limits the request to 10K hits per month for non-commercial licenses. We use the service from McAfee since it addresses both of the issues.}
These classifiers are able to classify URLs into 104 different fine-grained industry categories. 
Later in the paper, we discuss the categories that were most influential in deciding a resilience strategy.

The service from McAfee also provides a categorization based on reputation score (RS) of individual domain. RS indicates the level of risk a domain poses to a user's network, computer, or information~\cite{noauthor_mcafee_trustedsource}.
McAfee divides URLs into 4 \textit{Risk Levels} according to their RS from low to high: \textit{Minimal Risk (RS$<15$)}, \textit{Unverified ($15\leq$RS$<30$)}, \textit{Medium Risk ($30\leq$RS$<50$)}, and \textit{High Risk (RS$\geq50$)}.

Category labels from the URL categorization service indicate the functionalities of the domains based on their content.
When used outside of the context, the domains with certain characteristics
might lead to unintended consequences for the client's network and in turn, have  negative impact on businesses or individuals. 
For example, if all employees of a company with a limited bandwidth start using a streaming service, it may deteriorate the quality of network services within this company.
McAfee classifies URL categories into \textit{Risk Groups} based on the main aspect that is most at risk when this unintended situation occurs.\footnote{A more detailed description of these classifications is in \url{https://trustedsource.org/download/ts_wd_reference_guide.pdf}}
There are 7 risk groups that comprise of \textit{domains that pose a risk on}: 
\begin{description}
\item[Productivity] if they might reduce one's productivity at work, such as social media.
\item[Information] as they allow users to access work-irrelevant information.
\item[Liability]  if accessing them might be criminal, such as pirated contents.
\item[Security] of victimizing the users, such as phishing sites.
\item[Propriety] of serving non-criminal adult-only content, such as sexual material.
\item[Bandwidth] of providing bandwidth-consuming web pages, such as video streaming.
\item[Communication] as they allow direct communication with other users.
\end{description}
\subsection{Key Concepts and Methodology}
\label{sec:key_concepts}


\begin{table}
    \label{tab:possible_reactions}
    \caption{All possible reactions (R0 -- R3) for Dyn customers. \textit{Initial} shows domain status prior to the attack (Oct. 20, 2016), \textit{First Change} shows the first observed status change, and \textit{Final Change} shows the last.}
    \setlength{\tabcolsep}{2pt}
    \renewcommand{\arraystretch}{1}
    \centering
    \resizebox{\linewidth{}}{!}{
        \begin{tabular}{lccc}
            \toprule{}
            \multirow{2}{*}{\textbf{Reaction}} & \multicolumn{3}{c}{\textbf{Domain Status}}                                   \\ 
            \cmidrule(lr){2-4}
                                               & \textbf{Initial} & \textbf{First Change} & \textbf{Final Change}             \\ 
            \midrule{}
            \textbf{R0:\ }Do nothing                         & ex        &      ex                 & ex                  \\
                                               & non-ex    &             non-ex          & non-ex              \\
            \textbf{R1:\ }Become non-exclusive               & ex        & non-ex         & ex or non-ex\textsuperscript{*} \\
            \textbf{R2:\ }Unsubscribe and return     & ex or non-ex & unsub & ex or non-ex\textsuperscript{*} \\
            \textbf{R3:\ }Unsubscribe and not neturn & ex or non-ex & unsub & unsub               \\ 
            \bottomrule
            \multicolumn{4}{l}{\textit{ex: exclusive; non-ex: non-exclusive; unsub: unsubscribed.
            }} \\
            \multicolumn{4}{l}{\textit{\textsuperscript{*}We do not distinguish final domain statuses in these categories.}} \\
            \hline
        \end{tabular}
    }
\end{table}

We track the resilience strategy adopted by the domains that were customers of Dyn one day prior to the attack to analyze the change in relationship between Dyn and its customers as a consequence of the attack. We define two key concepts to systematically evaluate the changes in their approach to using managed DNS services:  \textit{Domain Status} and \textit{Domain Movement}. We define the \textit{reactions} we observe of these domains based on these two concepts.

\textit{Domain Status} shows the type of relationship between Dyn and a domain.  We identify the \textit{Domain Status} for a domain on a certain day from OpenINTEL data by looking at the presence of Dyn name servers (\texttt{*.dynect.*}) in the domain \texttt{NS} records on that day. \textit{Domain Status} for a customer of Dyn on a given day can either be:

\begin{description}
    \item[Exclusive domain] Domain status that indicates the use of \textit{only} name servers from Dyn.
    \item[Non-exclusive domain] Domain status that indicates the use of name servers from Dyn \textit{and at least one other} than Dyn.
    \item[Unsubscribed] This status represents that the domain used Dyn name servers one day prior to the attack but stopped using them on or before a certain day. For example, if \texttt{example.com} was a customer of Dyn one day before the attack, but stopped using its name servers 4 days after the attack then it would have a status of \textit{Unsubscribed} on and after that day, until it started using Dyn name servers again.
\end{description}

\textit{Domain Movement} captures the change in \textit{Domain Status} between two consecutive days. As Dyn allows users to list multiple name servers, \textit{Exclusive} domains can later become \textit{Non-exclusive} and vice versa. At the same time, a domain that is previously using Dyn services may decide to stop using its services and fully switch over to another provider, i.e., no longer shows Dyn name servers in \texttt{NS} records. We do not track domains that moved from being non-exclusive to exclusive as we are interested in domains that adopted a more resilient strategy (\cite{abhishta_measuring_2019} showed that a substantial number of domains did not change status from non-exclusive to exclusive during this period).

We attribute \textit{Domain Movement} based on the change in \textit{Domain Status} for a domain in the 20 days\footnote{We limit the analysis window to 20 days because after that news broke of Dyn being taken over by Oracle was, and hence we cannot attribute any change in Domain Status after 20 days solely to the attack.} after the attack as compared to the Domain Status one day prior to the attack.\footnote{We do not include the day of the attack because the attack may have overlapped with part of the DNS measurement.} We define the following movements, in order to capture the transitions between domain statuses:

\begin{description}
    \item[Moving to non-exclusive] Domains that changed the \textit{Domain Status} from exclusive to non-exclusive, i.e., use redundant services.
    \item[Unsubscribe] Domain response following the attack was to 
    stop using Dyn DNS service.
    \item[Return] Domains that changed the status from Unsubscribed to Exclusive or Non-exclusive domain, i.e.\ again became Dyn customers.
    
\end{description}


Based on the Domain Movement associated with each domain that was a customer of Dyn one day prior to the attack, we attribute the following reactions to the domain:

\begin{description}
    \item[Do nothing] Domains which \textit{did nothing} in response to the attack and hence had the same \textit{Domain Status} for the 20 days after the attack as one day before the attack (\textbf{R0}).
    \item[Become non-exclusive] Exclusive Dyn customers one day before the attack whose \textit{Domain Status} changed to non-exclusive (\textbf{R1}) at least once within 20 days after the attack.
    \item[Unsubscribe and return] Domains that used Dyn name servers one day before the attack, but unsubscribed and later returned within 20 days after the attack (\textbf{R2}). 
    \item[Unsubscribe and not return] Domains that unsubscribed from Dyn services within 20 days after the attack but did not return as Dyn customers (\textbf{R3}).
\end{description}
In this paper, we refer to domains that took reaction \textbf{R0} as \textit{non-reactive domains} and domains that took reactions \textbf{R1}/ \textbf{R2}/\textbf{R3} as \textit{re-active} domains. Table~\ref{tab:possible_reactions} shows all possible reactions associated to the observed change in \textit{Domain Status}.

There is a possibility of noise when we interpret \textit{Domain Movement} based on OpenINTEL data. Specifically, inconsistent configuration between multiple authoritative for the same domain could lead to mis-identification. 
To account for this possibility, we consider only status changes that persist for three days or longer. Moreover, we define a domain's reaction as its first movement we observe within 20 days after the attack.

As indicated in Table~\ref{tab:possible_reactions}, we do not further distinguish status of exclusive domains after they became non-exclusive, as well as of those that returned after unsubscribing.
Recall that becoming an exclusive customer involves having no (back-up) servers at another provider. 
Meanwhile, unsubscribing from Dyn service may avoid the risk of having a second or third service interruption from the provider. 
To be non-exclusive is relatively the most resilient option by having DNS service redundancy, even though it costs more or may add privacy risks.\footnote{By using open public DNS.}

We categorize domains based on four characteristics, i.e., \textit{popularity, industry sector, risk level,} and \textit{risk group}, that may influence the decision of Dyn customers to use a more resilient strategy. To analyze this, we follow this step-wise approach:

\begin{description}
    \item[Step 1] We identify domains using Dyn one day prior to the attack based on OpenINTEL data.
    \item[Step 2] We map domain characteristics of the domains identified in \textbf{Step 1} based on McAfee URL categorization data. 
    We add historical Alexa ranks (median ranks of 90 days prior to the attack) of domains to measure domain popularity. 
    \item[Step 3] We evaluate the impact of website popularity on resilience strategy irrespective of the industry sector.
    \item[Step 4] We use a multivariate logistic regression model \cite{gill2019generalized} to distinguish among industry sectors that are statistically significant factor in determining \textit{Re-active Domains} and \textit{Non-reactive Domains}. We evaluate in more detail the resilience strategy of the 20 most statistically significant industry sectors.
    \item[Step 5] We analyze the differences in resilience strategies of Dyn customers categorized by risk level and risk group.
\end{description}

 \section{Result and Discussion}
\label{sec:Result_and_Discussion}

We identify nearly 168K labeled domains with 94 unique industry categories from the categorization process that were Dyn customers on the day prior to the DDoS attack. Here we present the results of our analysis.

\subsection{Categorization of Dyn Customers}

We query category labels and reputation scores of the initial set of domains using a publicly available URL categorization service provided by McAfee ``Real-Time Database"~\cite{vallina_mis-shapes_2020}. \textbf{$\sim$62\% of the domains were categorized by McAfee.} 

\begin{figure}[t]
    \centering
    \includegraphics[width=\linewidth]{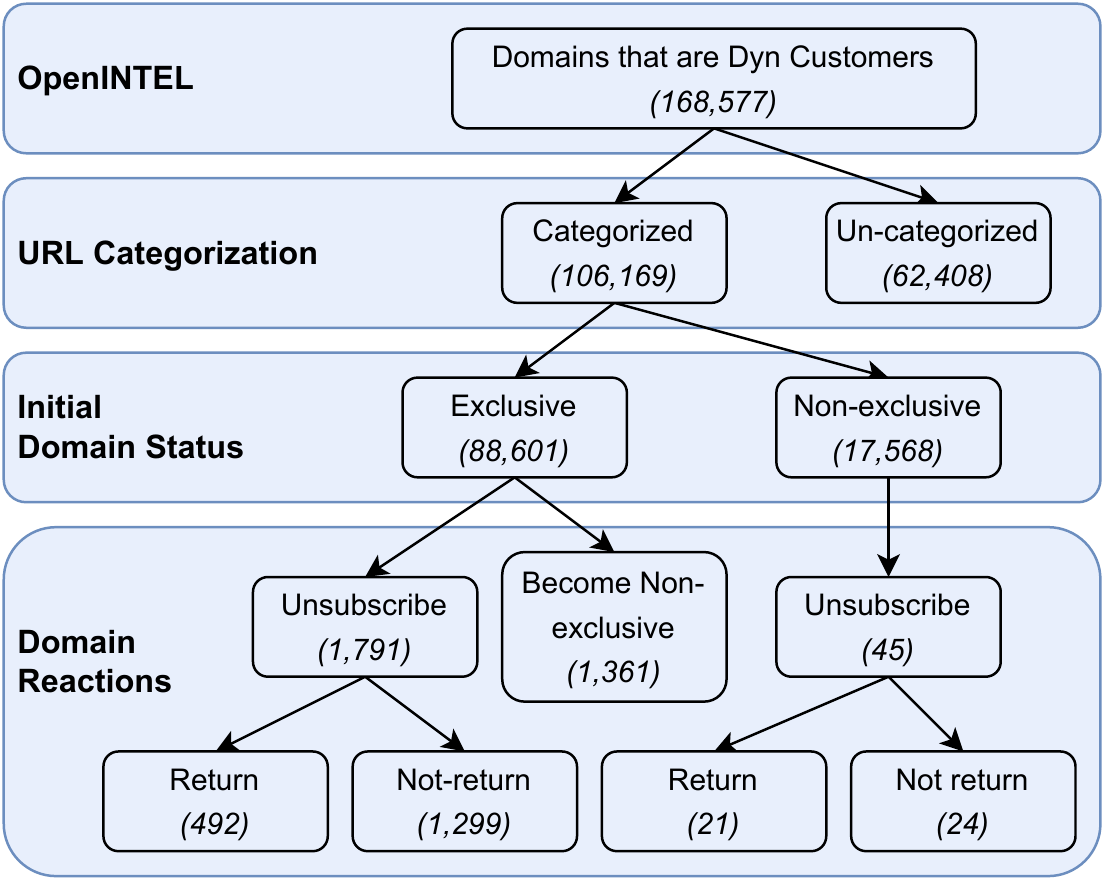}
    \caption{Dyn customer subdivision in category, domain status and reaction. We identify 169K customers and categorize them per our methodology. We distinguish exclusive and non-exclusive domains and subdivide them into reaction. $\sim$17K non-exclusive domains, 45 unsubscribed following the attack, and 24 out of 45 did not return to Dyn later.}
    \label{fig:macro_flowchart}
\end{figure}

Some domain names could not be categorized using McAfee; they are either no longer in operation or do not have enough features available for McAfee's classifier to make a deterministic classification. We refer to these domains as \textit{Un-categorized domains}. In this paper, we limit our analysis only to the domain names that McAfee successfully categorized, i.e., around 106K unique domains or nearly 62\% of all unique domains that used Dyn services one day prior to the attack (Fig.~\ref{fig:macro_flowchart}). 

Some domain names are mapped to multiple industry sectors and risk groups. We consider all retrieved category labels for our analysis. Hence, the total number of category labels retrieved for the domains exceeds the total number of unique domains.

We retrieve daily \textit{Domain Status} for all categorized domains from the measurement data for 20 days after the attack to observe status changes which we use to capture \textit{Domain Movements} as discussed in \S\ref{sec:methodology}.

\textbf{The number of \textit{Non-reactive Domains} is higher than \textit{Re-active Domains}.} The number of \textit{Non-reactive Domains} is significantly higher than those who became non-exclusive or unsubscribed (Fig.~\ref{fig:macro_flowchart}). We observed that more \textit{Exclusive Domains} either became non-exclusive or unsubscribed from Dyn services, as they had a greater probability of experiencing downtime due to the lack of redundancy in authoritative name servers.
Among non-exclusive domains, only 45 out of nearly 17K domains reacted. More than 85K out of $\sim$89K of \textit{Exclusive Domains} did not show any reaction (Fig.~\ref{fig:macro_flowchart}). This may be explained by the presence of cached \texttt{NS} records on DNS resolvers keeping several domains accessible during the outage~\cite{MouraIMC2018}. Also, for some organizations the short term cost of moving to another provider or adding redundancy may not justify benefits.

\textbf{The number of domains that became non-exclusive is not significantly different from the number of domains that unsubscribed.}
\textit{Re-active} domains showed two types of movement in terms of their decision to become non-exclusive or to unsubscribe from Dyn (Fig.~\ref{fig:macro_flowchart}). 
Among 3,197 re-active domains, 1,361 became non-exclusive and 1,836 unsubscribed.
There is no clear preference in the type of movement among domains. This observation motivates us to zoom into domains with similar characteristics, i.e., popularity or industry sectors, to discover specific preferences in the resilience strategy among similar Dyn customers.

\subsection{Popularity Influence on Decision to Re-act}
\label{sec:domain_popularity}

\begin{figure}
    \centering
    \includegraphics[width=\linewidth]{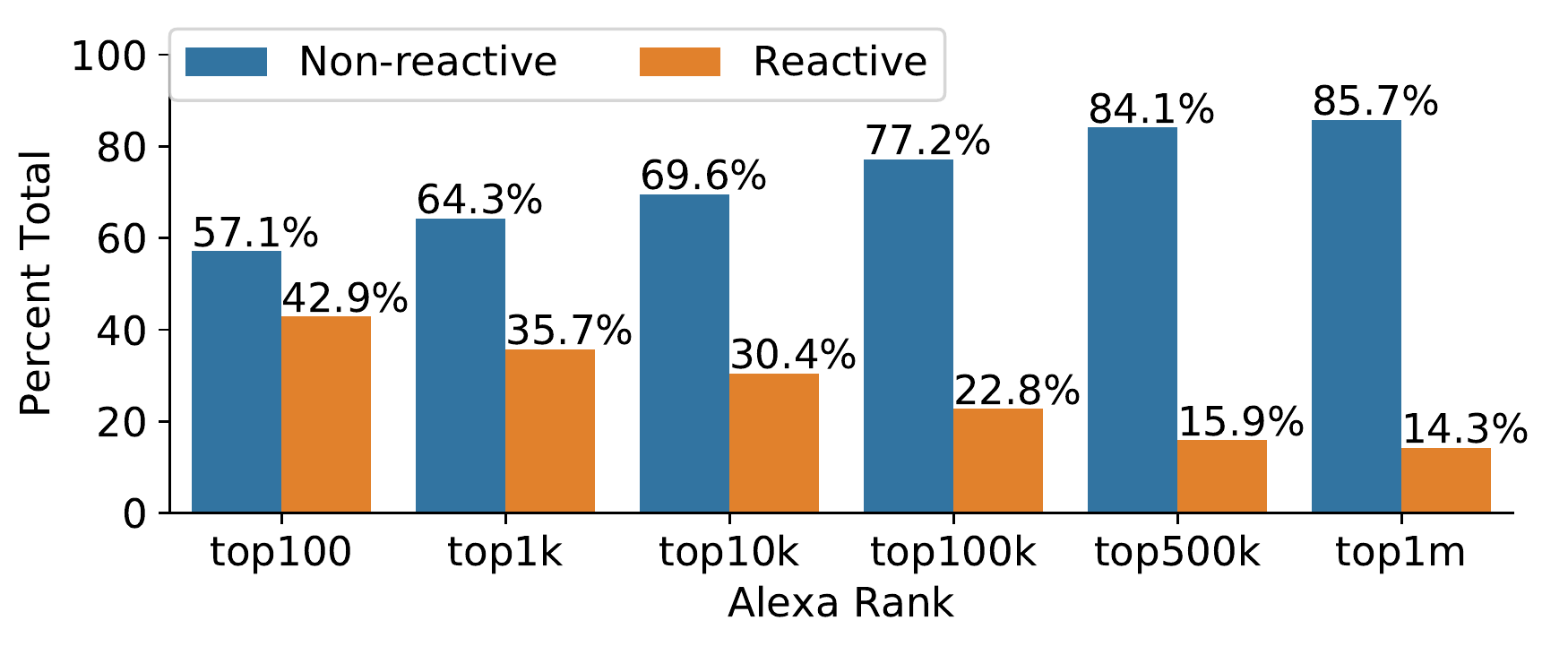}
    \caption{
        Reactions of Dyn customers -- re-active or non-reactive -- among different Alexa popularity groups. For example, among \textit{top100} Alexa customers, 42.9\% re-acted to the attack. (Note that larger groups are supersets, i.e., the \textit{top100} are included in the \textit{top1k})
    }
    \label{fig:by_popularity}
\end{figure}

We then incorporate Alexa rank data into our data set to indicate how popular each domain is. Out of $\sim$106K domains, we manage to retrieve popularity ranks of $\sim$5K domains that belong to the Top 1M Alexa list and ignore the ones which do not. We observe that domains with higher Alexa rank have a higher proportion of re-active domains (e.g., 42.9\% in top100) compared to those with lower rank (e.g., 14.3\% in top1m) as shown in Fig.~\ref{fig:by_popularity}:\ the percentage of re-active domains decreases as more domains with lower rank are included.
Popularity is thus one factor that appears to influence the decision to re-act.
A higher rank means higher traffic to these domains which may result in a higher revenue~\cite{truong_relationships_2018}.
Hence, owners of domains with higher rank will prefer minimizing downtime of their domains to avoid revenue loss. Meanwhile, owners of domains with lower rank may not be sensitive to short-term downtime.
Some reactions involve an extra cost, e.g., using multiple DNS service providers, which may not be worthwhile for businesses with smaller revenue.

\ 

\noindent
\textit{\textbf{Key takeaway}: We observe that customers running domains for more popular websites (based on Alexa rank) reacted to the attack with higher probability.}

\

\subsection{Industry Sector Influence on Re-action}
\label{sec:stat_analysis}

We discuss in more detail the Industry Sectors that had a significant impact on the resilience strategy adopted by the domain managers. 

To determine the impact of individual industry sectors on domain reactions, we model the resilience strategy adopted by the domains as a function of their industry sector using a multivariate logistic regression analysis. We model the binary decision of domains to re-act (re-act: \textbf{R1}/\textbf{R2}/\textbf{R3} or not react: \textbf{R0}) as a dependent variable (\ref{eq:logit_decision}). 

\begin{equation}
\label{eq:logit_decision}
    \text{\textit{logit}}(\pi_{b})=\log[\frac{\pi_{b}}{1-\pi_{b}}],
\end{equation}

\noindent where $\pi_{b} \in [0,1]$ represents the probability of a domain reacting and can be estimated using (\ref{eq_logit_estimation}).

\begin{equation}
    \label{eq_logit_estimation}
    \pi_{b}=\frac{\beta_{0}+\sum_{i}^{}\beta_{i}x_{i}}{1+(\beta_{0}+\sum_{i}^{}\beta_{i}x_{i})},
\end{equation}

\noindent where $x_{i} (i=1,2,3,....,94)$ represents each industry sector as a separate independent variable. $x_{i}$ has a value of 1 if a domain belongs to this industry sector, otherwise it has a value of 0. $\beta_{i}$ is the odds ratio and signifies the strength of correlation between this independent variable and the dependent variable. To implement logistic regression, we use \texttt{statsmodels} package in python~ \cite{seabold2010statsmodels}.

The regression model outputs three main metrics to indicate the influence of each category (shown in Table~\ref{tab:stat_result}).
\textit{Coefficient} ($\beta_{i}$) is used to calculate the factor by which the probability of a domain re-acting to the attack increases, if this domain belongs to Industry Sector $i$,  i.e., if the regression coefficient is estimated as $\beta_{i}$ then the probability of post-attack domain re-action increases by $e^{\beta_i}$ times as compared to the base case.

\begin{table}[t]
    \centering
    \caption{GLM coefficients for domains showing most prominent industry sectors that had a statistically significant influence on domain reactions (ordered by the number of domains in each industry sector).
    }
    \label{tab:stat_result}
    
   \begin{tabular}{lrr}
\toprule
Industry Sector&    Coefficient\textsuperscript{*} &       z-value \\
\midrule
Internet Services                    &  0.47 &   6.80 \\
Travel                               &  1.25 &  14.61 \\
Content Server                       &  1.26 &  11.00 \\
Motor Vehicles                       &  0.78 &   6.26 \\
General News                         &  2.17 &  24.05 \\
Job Search                           &  1.30 &   9.40 \\
Streaming Media                      &  2.15 &   6.57 \\
Gambling                             &  1.60 &   8.97 \\
Portal Sites                         &  1.74 &   6.51 \\
PUPs (potentially unwanted programs) &  2.83 &  12.28 \\
Malicious Sites                      &  2.61 &   8.91 \\
Social Networking                    &  2.47 &   7.75 \\
Personal Pages                       &  1.87 &   5.37 \\
Provocative Attire                   &  2.71 &   4.86 \\
For Kids                             &  3.07 &   4.93 \\
Interactive Web Applications         &  2.61 &   7.97 \\
Weapons                              &  1.98 &   5.32 \\
Professional Networking              &  3.58 &   9.34 \\
Visual Search Engine                 &  4.12 &   7.42 \\
Instant Messaging                    &  4.03 &   5.32 \\
        \bottomrule
        \textit{\textsuperscript{*}p$<$0.001}
    \end{tabular}
\end{table}

For example, domains in the \textit{Professional Networking} with a coefficient value 3.58 are 16 times (i.e.\ $\dfrac{e^{3.58}}{e^{0.78}}$) more likely to re-act than domains in the \textit{Motor Vehicles}, such as web sites of car manufacturers and sales, with a coefficient value 0.78.
Hence, we expect the proportion of re-active domains is higher in the former than the latter.
\textit{p-value} indicates if the coefficient of a domain is statistically significant or not. 
\textit{z-value} is calculated as the number of standard deviations the average decision of a certain industry sector is away from the average decision of the whole data set~\cite{mcleod_z-score_2019}. It indicates the reliability of the coefficient computed by the logistic regression model.

\begin{figure}[t]
    \centering
    \includegraphics[width=\linewidth]{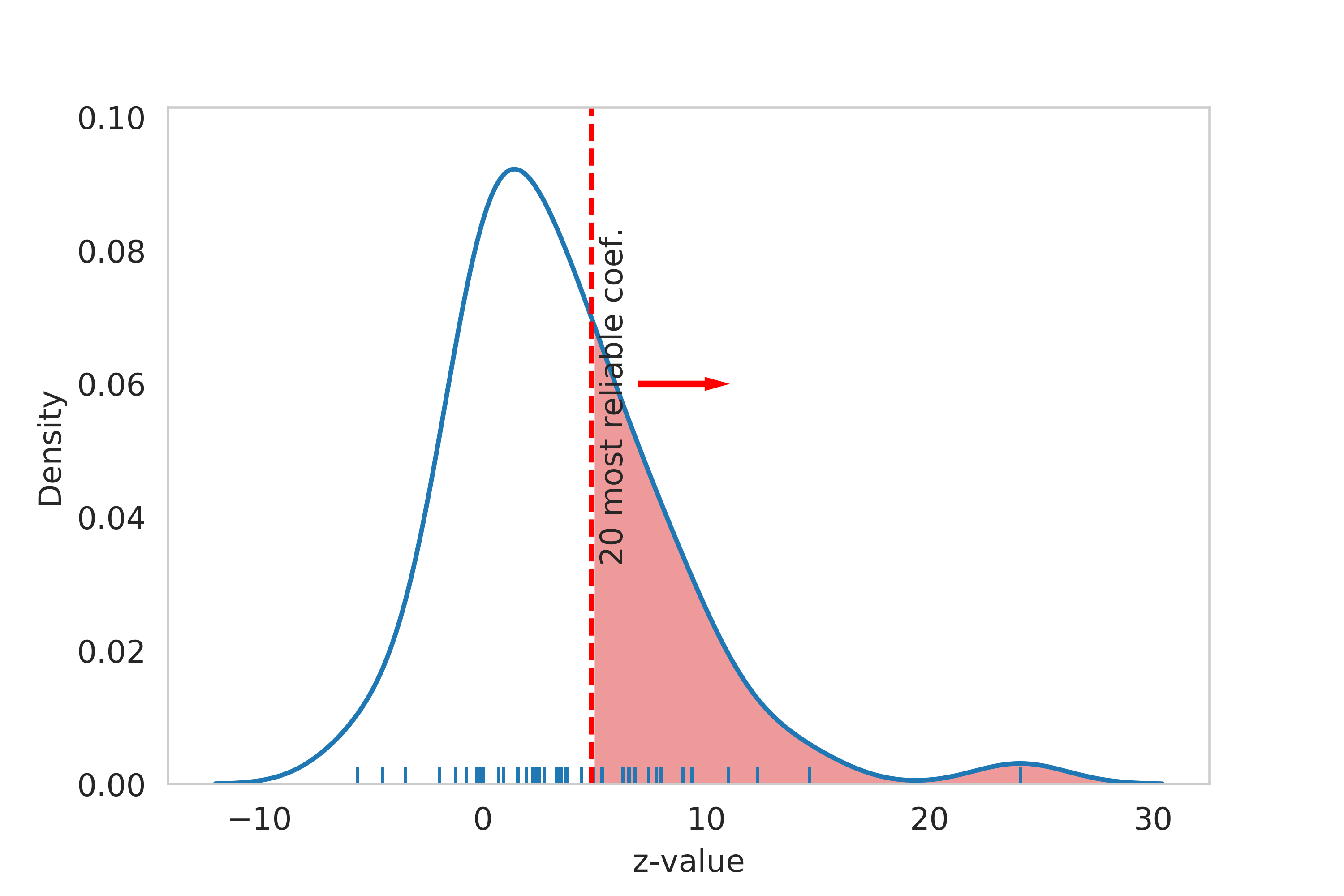}
    \caption{Distribution of z-values (signifies reliability) for industry sector coefficients in the linear regression model. We discuss in more detail the reactions of domains that belong to 20 industry sectors with the largest z-values, indicated by the shaded area.
    }
    \label{fig:z_hist}
\end{figure}

We discuss the type of reactions observed for domains that belong to any of the top 20 industry sectors whose coefficient was both highly reliable (greater than the median \textit{z-value} as shown in Fig.~\ref{fig:z_hist}) and statistically significant (\textit{p-value} $<0.05$).

The results indicate that some categories have more significant correlation to the resilience strategy than others.
Domains in industry sectors with higher coefficient have a higher probability to re-act following outages, such as \textit{General News} and \textit{Streaming Media}. 
DNS providers should anticipate the reaction from customers in these industry sectors more than other sectors.

\ 

\noindent
\textit{\textbf{Key takeaway:} By using a multivariate logistic regression model, we were able to calculate and reveal that for 34 out of 94 industry sectors, there is statistically significant influence on the post-attack reaction.}

\

We also observe a number of industry sectors which \textbf{do not} have statistically significant influence on the decision to re-act such as \textit{Religion}, \textit{Ideologies}, and \textit{Politics}.
Religion and Ideologies sectors include web pages that discuss spirituality and beliefs while Politics domains host discussions of political parties, views, and opinions. We suspect that domains in these sectors do not require high resilience.

\ 

\noindent
\textit{\textbf{Key takeaway:} Some industry sectors do not seem to have a statistically significant influence on the decision to re-act after the attack.}

\

\begin{figure*}[]
    \centering
    \includegraphics[width=\linewidth]{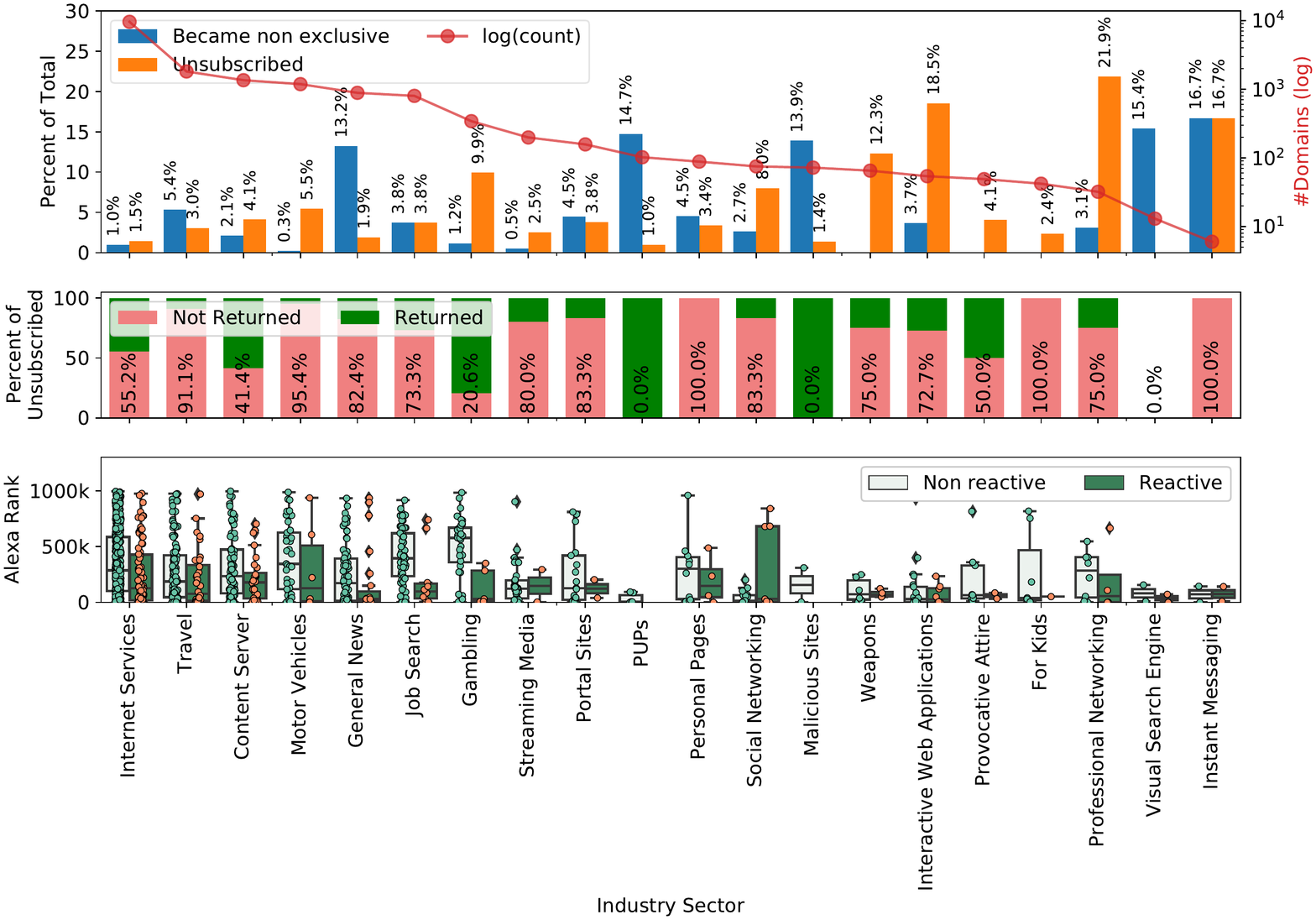}
    \caption{
        Behaviors of domains in selected industry sectors.
        \textbf{Top Plot} shows the proportions of domains with different reactions in each sector along with the sector sizes.
        \textbf{Middle Plot} shows the proportions of unsubscribed domains that did not return.
        \textbf{Bottom Plot} shows the distribution of Alexa ranks among domains that belong to a certain industry sector.
    }
    \label{fig:reactions_by_category}
\end{figure*}

We zoom into a number of significantly influential industry sectors (Table~\ref{tab:stat_result}).
Fig.~\ref{fig:reactions_by_category} shows how domains in these industry sectors reacted.
The \textit{top plot} shows proportions of re-active domains in each industry sector, characterized by their decision to become non-exclusive or to unsubscribe after the attack.
The diminishing line in the top plot indicates the size, i.e., number of domains, of each industry sector on a $log_{10}$ scale.
We observe that sector size does not directly correlate with the proportion of re-active domains in the sector.
However, we demonstrate that reactions vary among industry sectors in two ways: the proportion of \textit{re-active} domains and the dominant type of reactions.

From the \textit{top plot} in Fig.~\ref{fig:reactions_by_category}, we show that most domains in some industry sectors \textit{became non-exclusive} following the incident, i.e., used additional DNS providers, such as domains in \textit{General News}, \textit{Potentially Unwanted Programs (PUPs)}, \textit{Malicious Sites}, and \textit{Visual Search Engine}.
Domains in \textit{PUPs} and \textit{Malicious Sites} categories are made with an intention to infect user's computer with software or other malicious activity, e.g., Trojan horses and viruses.
Domains in \textit{Visual Search Engine} include web pages that provide searching functionality with image-specific results.
The reaction to use multiple DNS services might indicate that these domains have a high -- most probably financial -- value for the owners, hence, it is important to keep them up at all times.
After the attack, these domains used back-up services to be more resilient against future outages. Therefore, we assume that the revenue from the domains is large enough to compensate for the cost of using an additional DNS service, if present.

Meanwhile, domains in some industry sectors showed a significant preference to \textbf{unsubscribe}, i.e., they stopped using Dyn service, after the attack (Fig.~\ref{fig:reactions_by_category}).
Among them, we observe domains intended for adult audiences -- \textit{Gambling} sites, web pages about \textit{Weapons}, and sites showing \textit{Provocative Attires} -- and the domains intended for more general audiences such as \textit{Motor Vehicles} sites, \textit{Streaming Media}, and \textit{Social Networking} sites.
The reaction to unsubscribe after the attack might also indicate that domain owners value their domains and find it necessary to re-act after the attack.
However, we suspect that the cost of using a secondary DNS service is not worth the revenue from the domain, or the owner considers Dyn no longer reliable.

\

\noindent
\textit{\textbf{Key takeaway:} We see a difference in the type of reaction based on the industry sector. In some sectors, domains for the most part became non-exclusive following the attack (e.g., \textit{General News}), while for others, customers largely unsubscribed (e.g., \textit{Gambling}).}

\ 

The \textit{middle plot} shows the proportion of return among domains that unsubscribed in each sector.
This metric is important to measure the business impact of an outage incident as it indicates how many customers the provider loses for a longer term.
In some industry sectors such as \textit{Travel} (e.g., ticketing sites), \textit{Motor Vehicles}, and \textit{Personal Pages} (e.g., blogs), more domains \textbf{did not return} than those did return.
On the contrary, the behavior was quite the opposite in \textit{Gambling}, \textit{PUPs}, and \textit{Malicious Sites} sectors, i.e., most unsubscribed domains \textbf{returned} later.
In \textit{Visual Search Engine} sector, there is no domain that unsubscribed, hence, no data is shown in the plot for that sector.

\ 

\noindent
\textit{\textbf{Key takeaway:} In some industry sectors, domains that unsubscribed mostly did not return (e.g., \textit{Travel}), while in others, a higher proportion of domains unsubscribed and returned (e.g., \textit{Gambling}). This suggests that the industry sector affects decision-making.}


\ 

The \textit{bottom plot} in Fig.~\ref{fig:reactions_by_category} shows the distribution of Alexa ranks of the domain names -- where present -- within each industry sector categorized by their responses after the attack i.e., re-active or non-reactive. 
This plot reveals insights into the influence of popularity on the choice of resilience strategy adopted by domains in each industry sector.
%
For most industry sectors shown in Fig.~\ref{fig:reactions_by_category}, the average popularity of \textit{re-active} domains was higher than the average popularity of non-reactive domains. This is consistent with our previous findings in \S\ref{sec:domain_popularity} that higher popularity influences the domain decision to react.
However, we see that popularity was a more significant factor for some industry sectors than the others. \textbf{Non-overlapping inter quartile ranges (IQR)} for the box plots in \textit{Job Search} and \textit{Gambling} sectors indicate a strong correlation between popularity and the decision to react.
Meanwhile, in some industry sectors such as \textit{Internet Service}, domain popularity does not seem to have a strong correlation with the decisions to react, considering the substantial \textbf{overlap in the IQR} for the box plots of re-active and non-reactive domains.

\ 

\noindent\textit{\textbf{Key takeaway:} In some industry sectors (e.g., Job Search), the influence of popularity on the decision to re-act is stronger than in others (e.g., Internet Service).}

\subsection{Risk Level Influence on Decision to React}
\label{sec:domain_risk_level}



\begin{figure*}[t]
    \centering
    \includegraphics[width=\linewidth]{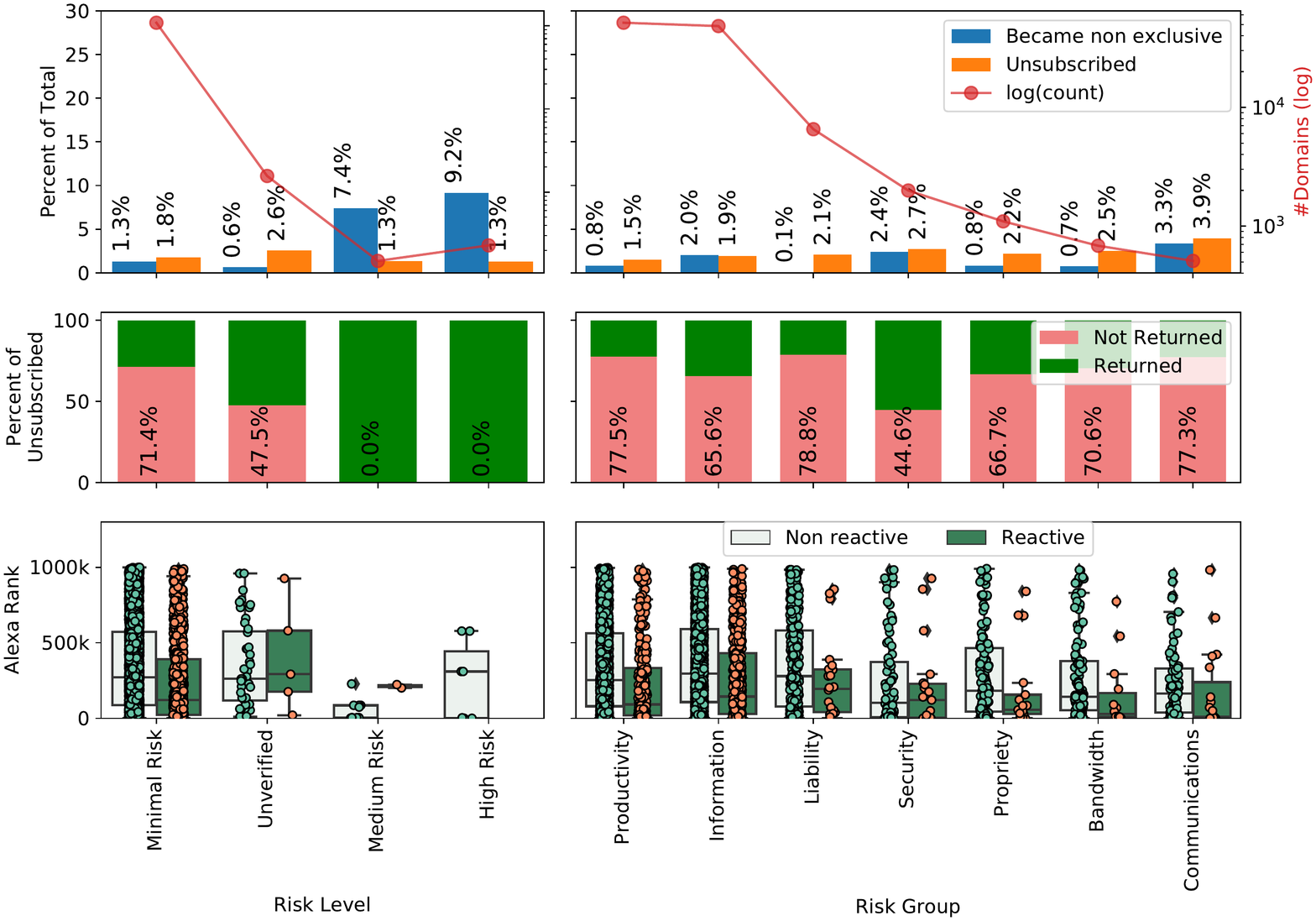}
    \caption{
    Behaviors of domains with different \textit{Risk Levels} and \textit{Risk Groups}. 
    \textbf{Top Plot} shows the proportions of domains with different reactions in each group along with the group sizes. 
    \textbf{Middle Plot} shows the proportions of unsubscribed domains that did not return. 
    \textbf{Bottom Plot} shows the comparison in the distribution of Alexa ranks of re-active and non-reactive domains within each category.
    }
    \label{fig:reaction_by_risk_type}
\end{figure*}

We analyze how risk level of the domains influences their reaction following the attack. Fig.~\ref{fig:reaction_by_risk_type} shows the behavior of domains with different risk categories, characterized by risk level and risk group.
Recall that McAfee categorizes domains based on the reputation score into four risk levels as shown in \S\ref{sec:methodology}.
Most domains using Dyn services are categorized as `Minimal Risk'. Out of 106K domains of Dyn customers, 104K are in Minimal Risk while domains in Unverified, Medium Risk, and High Risk categories only account for 2K domains (Fig.~\ref{fig:reaction_by_risk_type}).

Some domains with higher risk levels also need high availability, hence, using the managed DNS service from Dyn.
As illustrated by the \textit{top plot} in Fig.~\ref{fig:reaction_by_risk_type}, some Dyn customers are domains in \textit{medium} and \textit{high risk} levels and they show a similar preference to become non-exclusive after the attack, i.e., use redundant DNS services.
Moreover, most domains with medium and high risks returned if they unsubscribed, (the \textit{middle plot}).
We also observed this behavior among domains intended to victimize the users, e.g., PUPs and Malicious Sites (\S\ref{sec:stat_analysis}).
%

We see a significant \textbf{overlap in the IQR} in the box plots of Alexa ranks for domains categorized as minimal risk (see:\ \textit{bottom plot} of Fig.~\ref{fig:reaction_by_risk_type}). This shows that popularity (for this aggregate set of domains) does not play a major role in deciding the post-attack reactions. The number of re-active domains categorized as unverified or medium risk are too low to derive any conclusions about the influence of popularity on the observed reaction. Meanwhile, no domain categorized as a high risk belonged to Alexa Top 1M list. This observation is again consistent with our result in \S\ref{sec:stat_analysis} with domains categorized as \textit{malicious domains}.


\

\noindent
\textit{\textbf{Key takeaway:} Customers with higher risk levels mostly became non-exclusive after the attack and the majority of those who unsubscribed instead also later returned.} 

\ 

In addition, considering the \textit{top} and \textit{middle plots} in Fig.~\ref{fig:reaction_by_risk_type}, we observe that majority of the domains with minimal risk stopped using Dyn services permanently following the attack, i.e.\ nearly 1K domains or 71.4\% of those that unsubscribed did not return.
This shows that once customers of DNS service providers unsubscribe there is less chance for them to re-subscribe to the service. Other DNS providers may use similar heuristics to estimate customer loss following an outage.

\ 

\noindent
\textit{\textbf{Key takeaway:} Our results show that if domains in the minimal risk category unsubscribe, the probability that they stay away for a longer time is higher.}
    
\subsection{
Risk Group Influence on Decision to React}
\label{sec:domain_risk_type}



Fig.~\ref{fig:reaction_by_risk_type} shows the re-actions among domains that belong to different Risk Groups.
According to the \textit{top plot}, domains with \textit{Liability}, \textit{Bandwidth}, and \textit{Propriety} risks show a slight preference to unsubscribe after the attack. More than 60\% of the unsubscribed domains in these categories did not return. 

Meanwhile, as shown in the \textit{middle plot}, we observe similar return behaviors of domains in different risk groups, i.e., more than half of them did not return after they unsubscribed, except for domains in \textit{Security} risk group.


As shown in the \textit{bottom plot} of Fig.~\ref{fig:reaction_by_risk_type}, the popularity of re-active domains in all risk groups (except Security) is higher than the popularity of the domains that do not re-act. For 5 of the 7 risk groups, we observe that the re-active domains are concentrated in relatively higher Alexa ranks. This indicates that for these domains (i.e.\ \textit{Liability, Security, Propriety, Bandwidth}, and \textit{Communications}), popularity played an important role in the decision of domain owners to re-act. Interestingly, these domains also include websites providing streaming and messaging services.    


This result is consistent with our observation in \S\ref{sec:domain_popularity} that higher popularity shows a correlation with the decision to re-act after the attack.
In \textit{Productivity} and \textit{Information} risk groups (eg.\ social media websites), we see a substantial \textbf{overlap in the IQR} of the re-active and non-reactive box plots, which indicates a weaker correlation of the domain popularity with the decision to react.

%

\

\noindent\textit{\textbf{Key takeaway:} Domains in some risk groups (e.g., Liability) mostly unsubscribed after the attack.}

\

\subsection{Analysis on Domain Autonomous Systems}
\label{sec:domain_as}

A single party can operate multiple domains that rely on Dyn for DNS. For example, Alphabet Inc. operates, among others, google.com and gmail.com. So far we have not considered the potential effects of a single party making decisions (i.e., driving domain reactions) for multiple domains. 
As a first step towards exploring if multiple re-actions can be correlated with decisions made by a single party, we measure the variety of reactions of Dyn customers within each Autonomous System (AS).

\begin{figure}[t]
    \centering
    \includegraphics[width=\linewidth]{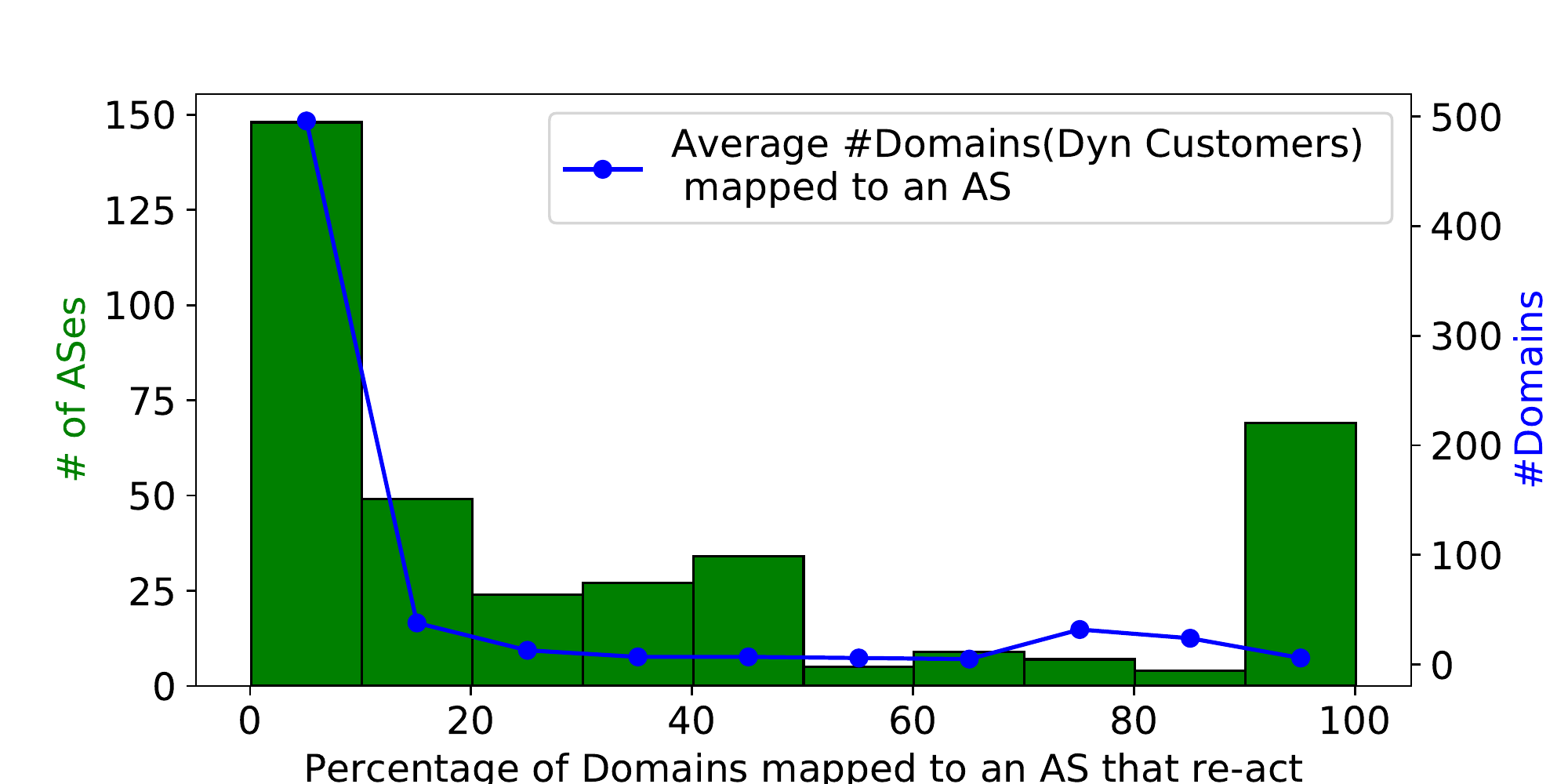}
    \caption{Distribution of variety in re-actions among Domains mapped to an AS.}
    \label{fig:AS_Size_Histogram}
\end{figure}

We associate domains to their hosting network at the time of the Dyn attack by mapping IP addresses (i.e., \texttt{A} records) to AS numbers, using IP-prefix-to-AS data from CAIDA~\cite{the_caida_ucsd_routeviews_2008}. 
We then calculate the percentage of re-active domains that re-acted in each AS. In other words, if two domains (both Dyn customers) mapped to a single AS and one re-acted, the percentage of re-active domains for that AS is 50\%.
Fig.~\ref{fig:AS_Size_Histogram} shows the resulting distribution. We only include ASes that were mapped to at-least 1 re-active domain in this plot. Nearly 1.6K different ASes, each in which on average 14 domains were hosted, showed no re-active domains at all.
We observe that in the case of about 200 smaller ASes (one to only a few domains hosted), the domains all showed the same behavior (i.e., 100\% re-acted). This suggests that parties that operated several domains drove the reaction for multiple domains, but on the order of a few hundred domains in total based on our initial view. This approach of course does not account for cases where a single party hosts their domains in multiple networks, which we leave for future work.

\

\noindent\textit{\textbf{Key takeaway:} Domains mapped to the same AS showed different re-actions.}

\

\section{Conclusion and Implications}
\label{sec:conclusion}




Unplanned outages of managed DNS services can lead to financial losses for customers. Depending on how customers value online availability, outages could prompt customers to switch to a different DNS service provider or to add a (secondary) provider to add resilience.
We leveraged large-scale DNS measurement data to infer 168K Dyn customers at the time of the infamous 2016 DDoS attack and tracked the choices made by customers following the attack. We also used a publicly available URL categorization service and website popularity metrics to characterize customers. We demonstrated the impact of factors such as domain popularity, industry sector, risk level, and risk group on the adopted resilience strategy.

\vspace{0.4em}
\noindent Our findings are as follows:
\begin{itemize}
    \item 
    We confirmed the intuition that domains with a higher popularity (i.e., larger user base) are more likely to re-act (40\% of customers on the Alexa Top 100 in contrast to only 15\% in the Top 500K);
    \item 
    We inferred 94 industry sectors for customers and revealed that 34 sectors statistically significantly influenced post-attack customer behavior; 
    \item 
    We unveiled that customers in different sectors adopted different resilience strategies. For example, in some sectors customers typically stopped using Dyn altogether whereas in others they added an additional DNS provider to reduce the risk of unavailability;
    
    \item  We showed that the popularity of domain names is more tightly related to post-attack behaviors in some industry sectors over others;
    \item 
    A large proportion of domains categorized as `minimal risk' that unsubscribed from Dyn, did not return (more than 71\% did not return). We observe that domains categorized as medium or high risk prefer to become non-exclusive after the attack (9\% became non-exclusive while 1\% unsubscribed). 
    We also see that domains in some risk groups such as \textit{Liability} (e.g., pirated movie portals) showed a slight preference to unsubscribe after the attack,  with 0.1\% becoming non-exclusive and 2\% unsubscribed.
\end{itemize}

The results of our study could help managed DNS providers to estimate the impact of an unplanned outage based on their customer portfolio. Businesses such as cloud service providers could make use of a methodology similar to ours to understand the impact of unplanned downtime on their customer portfolio~\cite{Yasir2021}.

We show that \textit{domain popularity}, \textit{industry sector}, and -- to some extent -- \textit{risk groups} \& \textit{risk level}, are factors that can be used to understand the choice of resilience strategy. 

\subsection{Limitations and future research}
The nature of our data sources and assumptions that we need to make lead to a few limitations.
\textit{First}, the URL categorization service we use in this study might exclude domains without hosted content.
Recall that the domain content is a feature in our categorization process. 
Moreover, one should be aware that all existing URL categorization services, including McAfee's, have accuracy concerns~\cite{vallina_mis-shapes_2020}.
\comment{TBD To account for this, we randomly sample 100 domain names labeled by McAfee to manually verify the given labels and yield that XX\% of them are correctly classified.}
\textit{Second}, Alexa does not rank non-website domains (i.e., other networked services), despite their high traffic. Moreover, Alexa ranks the Top 1 million most popular websites only. Consequentially, most Dyn customers that we infer ($\sim$163K out of $\sim$168K) do not have an associated rank.
In addition, Alexa Top 1M is prone to bias and a rigorous alternative exists, i.e., Tranco\cite{le_pochat_tranco_2019}. However, Tranco did not yet exist when the attack happened in 2016.
\textit{Last}, the insights we draw are based on the analysis of data connected to a single attack event. Hence, we understand that similar behavior may not be observed for another event. However, researchers could apply our methodology to other similar attacks and outages to improve the generalizability of our findings.  

As one area of future work, we plan to analyze the response following outages (attack related or otherwise) in other online services such as CDN or web hosting using a similar approach.
Even though these outages would also lead to downtime, the behavior observed may be different given that the operator/manager might need to employ greater effort to opt for a more resilient strategy. 
In addition, considering the development of the DNS ecosystem since 2016, we plan to analyze more recent incidents to gain updated insight into the topic and study what changed in the last six years.




\bibliographystyle{IEEEtran}
\bibliography{IEEEabrv,references.bib}

\comment{

\appendices

\section{Experimental data}
\lipsum[1-4]
\section{Theorem proofs}
\lipsum[5-6]

\vspace{12pt}
\color{red}
IEEE conference templates contain guidance text for composing and formatting conference papers. Please ensure that all template text is removed from your conference paper prior to submission to the conference. Failure to remove the template text from your paper may result in your paper not being published.
}

\end{document}